\documentclass[useAMS,usenatbib]{mn2e}
\input psfig.sty
\usepackage{dcolumn}% Align table columns on decimal point
\usepackage{bm}% bold math
\usepackage{color}
\usepackage{epsfig}
\usepackage{graphics}% Include figure files
\usepackage{dcolumn}% Align table columns on decimal point
\usepackage{bm}% bold math
\newcommand{\bq}{\begin{equation}}		%%
\newcommand{\eq}{\end{equation}}		%%
\newcommand{\bqn}{\begin{eqnarray}}		%%
\newcommand{\eqn}{\end{eqnarray}}		%%
\def \be {\begin{equation}} 
\def \ee {\end{equation}} 
\def \bea {\begin{eqnarray}} 
\def \eea {\end{eqnarray}} 

\begin{document}

\title[Cosmic homogeneity: a spectroscopic and model-independent measurement]{Cosmic homogeneity: a spectroscopic and model-independent measurement}

\author[R. S. Gon\c{c}alves et al.]% G. C. Carvalho, C. A. P. Bengaly Jr., J. C. Carvalho, A. Bernui, J. S. Alcaniz and R. Maartens]
{
 \parbox{\textwidth}{
  R. S. Gon\c{c}alves$^{1}$\thanks{E-mail: \texttt{rsousa@on.br}},
  G. C. Carvalho$^{1,2}$,
  C. A. P. Bengaly Jr.$^{1,3}$,
  J. C. Carvalho$^{1}$,
  A. Bernui$^{1}$,
  J. S. Alcaniz$^{1,4}$,
  and
  R. Maartens$^{3,5}$,
 }
 \vspace{0.4cm}\\
 \parbox{\textwidth}{
  $^{1}$Observat\'orio Nacional, 20921-400, Rio de Janeiro - RJ, Brasil\\
  $^{2}$Departamento de Astronomia, Universidade de S\~ao Paulo, 05508-090 S\~ao Paulo - SP, Brasil\\
  $^{3}$Department of Physics \& Astronomy, University of the Western Cape, Cape Town 7535, South Africa\\
  $^{4}$Physics Department, McGill University, Montreal QC, H3A 2T8, Canada\\
  $^{5}$Institute of Cosmology \& Gravitation, University of Portsmouth, Portsmouth PO1 3FX, United Kingdom\\
 }
}

\pagerange{\pageref{firstpage}--\pageref{lastpage}}

\pubyear{2017}

\maketitle

\label{firstpage}

\begin{abstract}
Cosmology relies on the Cosmological Principle, i.e., the hypothesis that the Universe is homogeneous and isotropic on large scales. This implies in particular that the counts of galaxies should approach a homogeneous scaling with volume at sufficiently large scales. Testing homogeneity is crucial to obtain a correct interpretation of the physical assumptions underlying the current cosmic acceleration and structure formation of the Universe. In this {\em Letter}, we use the Baryon Oscillation Spectroscopic Survey to make the first spectroscopic and model-independent measurements of the angular homogeneity scale $\theta_{\rm h}$. Applying four statistical estimators, we show that the angular distribution of galaxies in the range $0.46 < z < 0.62$ is consistent with homogeneity at large scales, and that $\theta_{\rm h}$ varies with redshift, indicating a smoother Universe in the past. These results are in agreement with the foundations of the standard cosmological paradigm.
\end{abstract}


%\tableofcontents

\section{Introduction}

The Cosmological Principle constitutes one of the most fundamental pillars of modern cosmology. In  past decades, it has been indirectly established as a plausible physical assumption, given the observational success of the standard $\Lambda$CDM cosmology, which assumes large-scale homogeneity and isotropy, with structure formation described via perturbations. Although isotropy has been directly tested~\citep{blake02, bernui14, tiwari16, bengaly16, schwarz16, bengaly17, bernal17, javanmardi17}, homogeneity is much harder to probe by observations (see, e.g.,~\citealt{clarksonmaartens10}, \citealt{maartens11},\citealt{clarkson12}). 

As is well known, the smaller the scale we observe, the clumpier the universe appears. However, non-uniformities such as groups and clusters of galaxies, voids, walls, and filaments, are expected in a Friedmann-Lema\^itre-Robertson-Walker (FLRW) universe according to cosmological simulations. In such a background, a transition scale is also expected, above which the patterns composed by these structures become smoother, eventually becoming indistinguishable from a random distribution of sources. This homogeneity scale $r_{\rm h}$ has been identified and estimated at $70 - 150 \, \mathrm{Mpc}/h$, using data from several galaxy and quasar surveys \citep{hogg05, scrimgeour12, nadathur13, alonso15, pandey15, laurent16, sarkar16, ntelis17}, although other authors have claimed no evidence for it \citep{syloslabini98, syloslabini11, park17}. In the context of the $\Lambda$CDM paradigm, an upper limit for the homogeneity scale was estimated by~\cite{yadav10} to be $r_{\rm h} \sim 260 \, \mathrm{Mpc}/h$. 

Tests of homogeneity of the matter distribution by counting sources in spheres or spherical caps are not direct tests of geometric homogeneity, i.e. of the Cosmological Principle. Source counts on spatial hypersurfaces inside the past lightcone cannot be accessed by this method, since the counts are restricted to the intersection of the past lightcone with the spatial hypersurfaces. Instead, source counts provide consistency tests: if the count data show that the matter distribution does not approach homogeneity on large scales, then this can falsify the Cosmological Principle. Alternatively, if observations confirm an approach to count homogeneity, then this strengthens the evidence for geometric homogeneity -- but cannot prove it. A test of homogeneity of the galaxy distribution that does probe inside the past lightcone has been developed by~\cite{heavens11, hoyle12} -- but this test is unable to determine a homogeneity scale. 

When a length scale $r_{\rm h}$ is used to probe homogeneity, a further assumption is made -- a fiducial FLRW model is assumed a priori, in order to convert redshifts and angles to distances. In order to circumvent this model dependence, one can use an angular homogeneity scale $\theta_{\rm h}$ \citep{alonso14}. It was shown by~\cite{alonso15} that the $\theta_{\rm h}$ determined from the 2MASS photometric catalogue is consistent with $\Lambda$CDM-based mock samples within $90\%$ confidence level.

In this {\it Letter}, we make tomographic measurements of $\theta_{\rm h}$ in the Luminous Red Galaxies (LRG) sample from the Baryon Oscillation Spectroscopic Survey (BOSS), data release DR12. Because DR12 is a dense, deep galaxy catalogue covering roughly 25\% of the sky, it provides an excellent probe of the large-scale galaxy distribution, allowing us to make robust measurements in six very thin ($\Delta z = 0.01$), separated redshift shells in the interval $0.46 < z < 0.62$. This also avoids the additional correlations that would arise due to projection effects~\citep{sarkar16,carvalho16,carvalho17}. To our knowledge, this is the first time that the characteristic homogeneity scale is obtained with a spectroscopic and model-independent measurement, at intermediate redshifts. In addition, we are able to determine the redshift evolution of $\theta_{\rm h}$. We ensure further robustness by using four different estimators, which produce results that are compatible with each other and with the 
predictions of 
standard 
cosmology, without assuming any cosmological model a priori.

\section{Analysis}

\subsection{Observational data}

The total effective area covered by BOSS DR12 is 9,329 deg$^2$, with completeness parameter $c > 0.7$. As in previous BOSS data releases, DR12 is divided into two target samples: LOWZ (galaxies up to $z \simeq 0.4$) and CMASS (massive galaxies with $0.4 < z < 0.7$). They cover different regions in the sky, named north and south galactic cap. Here we are interested in exploring the homogeneity transition at redshifts $z > 0.46$, and we use only the north galactic cap of the CMASS LRG sample. 

We divide the DR12 CMASS sample into six thin redshift bins of width $\Delta z = 0.01$, between $0.46 < z < 0.62$. As observed in Table \ref{table1}, the number of galaxies in each bin is $N_{\rm galaxies} \geq 18,800$, thus providing good statistical performance for the analysis. 
Moreover, we choose non-contiguous bins to suppress correlations between neighbouring bins. 

\subsection{Methodology}

For a homogeneous angular distribution, the number counts in spherical caps of angular radius $\theta$ are given by
\be\label{barn}
\bar N(\theta)=\bar n \, A(\theta), \quad A(\theta)=2\pi(1-\cos\theta),
\ee
where $\bar n$ is the angular number density and $A$ is the solid angle of the cap. 
If the observed number is $N$, we define the scaled number count ${\cal N}=N/\bar N$, which  is 
obtained in four different ways as presented below. The correlation dimension is
\begin{equation}
\label{defD2}
D_2(\theta) \equiv {d\ln N \over d\ln\theta}= {d\ln {\cal N} \over d\ln \theta} + {\theta\sin\theta \over 1-\cos\theta} ,
\end{equation}
where the second equality follows from Eq. (\ref{barn}). The homogeneous limit is 
\be\label{d2h}
D_{2{\rm h}}(\theta)= {\theta\sin\theta \over 1-\cos\theta}\simeq 2,
\ee 
where the approximation is accurate to sub-percent level for $\theta \leqslant 0.34 \; rad$, i.e., $\sim 20^\circ$.

Estimators for ${\cal N}$ are defined below, based on their counterparts for $r_{\rm h}$ \citep{alonso15,laurent16,ntelis17}. In order to estimate the observational results we need to compare the observational data, previously described, with mock catalogues. In our analysis we use twenty random catalogues, generated by a Poisson distribution with the same geometry and completeness as the SDSS-DR12 {\footnote{https://data.sdss.org/sas/dr12/boss/lss/}}. \\

%------------------------------------------------------------------------------------------------------------------------
\begin{table}
\centering
\begin{tabular}{|c|c|c|}
\hline
\,\,$\bar{z}$ \,\,& \,\,redshift bins \,\,& \,\,$N_{\rm galaxies}$\,\, \\
\hline
\,\,\,0.465 \,\,& 0.46 - 0.47 & 22551 \\
\,\,\,0.495 \,\,& 0.49 - 0.50 & 31763 \\
\,\,\,0.525 \,\,& 0.52 - 0.53 & 32794 \\
\,\,\,0.555 \,\,& 0.55 - 0.56 & 29486 \\
\,\,\,0.585 \,\,& 0.58 - 0.59 & 23997 \\
\,\,\,0.615 \,\,& 0.61 - 0.62 & 18800 \\
\hline
\end{tabular}
\caption{The six redshift bins used in the analysis and their properties: mean redshift, 
bin width, and number of galaxies.}
\label{table1}
\end{table}
%------------------------------------------------------------------------------------------------------------------------

\subsubsection{Average}

This is the most common approach in the literature \citep{alonso14,ntelis17}. We define a cap in the sky of a given angular separation $\theta$ around one galaxy, counting how many galaxies are  inside this region. We repeat the process considering each galaxy as the centre ('cen') of a cap for different angular separation values, and for each redshift bin, thus obtaining a number count average in each case. The same process is replicated for the random catalogue, and  we define the estimator as the ratio of the averages:
\begin{equation}
\label{NMean}
{\cal N}(\!<\!\theta)^{\rm Ave} \equiv \frac{  \sum_i N^{\rm obs}_{i\,{\rm cen}}/{M^{\rm obs}_{\rm cen}}}{ 
\sum_i N^{\rm ran}_{i\,{\rm cen}} /{M^{\rm ran}_{\rm cen}}} \, ,
\end{equation}
where the total number of galaxies used as centres of caps are equal in both  catalogs, $M^{\rm obs}_{\rm cen} = M^{\rm ran}_{\rm cen}$. Then we calculate $D_2(\theta)^{\rm Ave}$ via Eq. (\ref{defD2}). Finally, we repeat the previous steps for twenty random catalogues, obtaining a mean value and a standard deviation for $D_2(\theta)^{\rm Ave}$.\\

\noindent

\subsubsection{Centre}

First we calculate the ratio of the observed and random counts-in-caps centred on the first galaxy, using the equivalent position in the random catalogue. Then we repeat the process for each centre in both catalogues, obtaining
\begin{equation}
\label{Ncenter}
{\cal N}(\!<\!\theta)^{\rm Cen} \equiv 
\frac{1}{M^{\rm ran}_{\rm cen}} \sum \frac{N^{\rm obs}_{i\,{\rm cen}}}{N^{\rm ran}_{i\,{\rm cen}}} \, .
\end{equation}
We calculate $D_2(\theta)^{\rm Cen}$ via Eq. (\ref{defD2}), and then repeat the previous steps for twenty random datasets in order to calculate its mean and standard deviation. \\

\noindent

\subsubsection{Peebles-Hauser (PH)}

We follow the Peebles-Hauser \citep{peebleshauser74} estimator, but instead of using the number of galaxies, we estimate the scaled counts-in-caps by the number of pairs within a given angular separation in the catalog. We define $DD(\theta)$ as the number of pairs of galaxies (for a given $\theta$) normalized to the total number of pairs, $M^{\rm obs}(M^{\rm obs} - 1) / 2$. We define $RR(\theta)$ equivalently for the random catalogue. Then
\begin{equation}
\label{NPH}
{\cal N}(\!<\!\theta)^{{\rm PH}}\! \equiv \!
\frac{\sum^{\theta}_{\phi = 0} DD(\phi)}{\sum^{\theta}_{\phi = 0} RR(\phi)} \, ,
\end{equation}
and $D_2(\theta)^{\rm PH}$ follows from Eq. (\ref{defD2}). As above, this procedure is repeated for the other random catalogues, from which obtain the mean and standard deviation for $D_2(\theta)^{{\rm PH}}$. \\

\noindent

\subsubsection{Landy-Szalay (LS)}

We use an estimator based on the Landy-Szalay correlation function \citep{landyszalay93}. In addition to the previous definition, we  define 
$DR(\theta)$ as the number of pairs of galaxies between the observational and random catalogues, for a given $\theta$, normalized by $M^{\rm obs} \, M^{\rm ran}$. Following a similar routine to the PH estimator, we obtain
\begin{equation}
\label{NLS}
{\cal N}(\!<\!\theta)^{{\rm LS}} \equiv 
1 + \frac{\sum^{\theta}_{\phi = 0} [DD(\phi)-2DR(\phi)+RR(\phi)]}{\sum^{\theta}_{\phi = 0} RR(\phi)} .
\end{equation}
We again calculate $D_2(\theta)^{\rm LS}$ via Eq. (\ref{defD2}), and after repeating this step for the other random data, we obtain the mean and standard deviation for $D_2(\theta)^{{\rm LS}}$. 

\subsubsection{Estimation of $\theta_{\rm h}$}

\begin{figure}
\vspace{-0.5cm}
\includegraphics[scale=0.45]{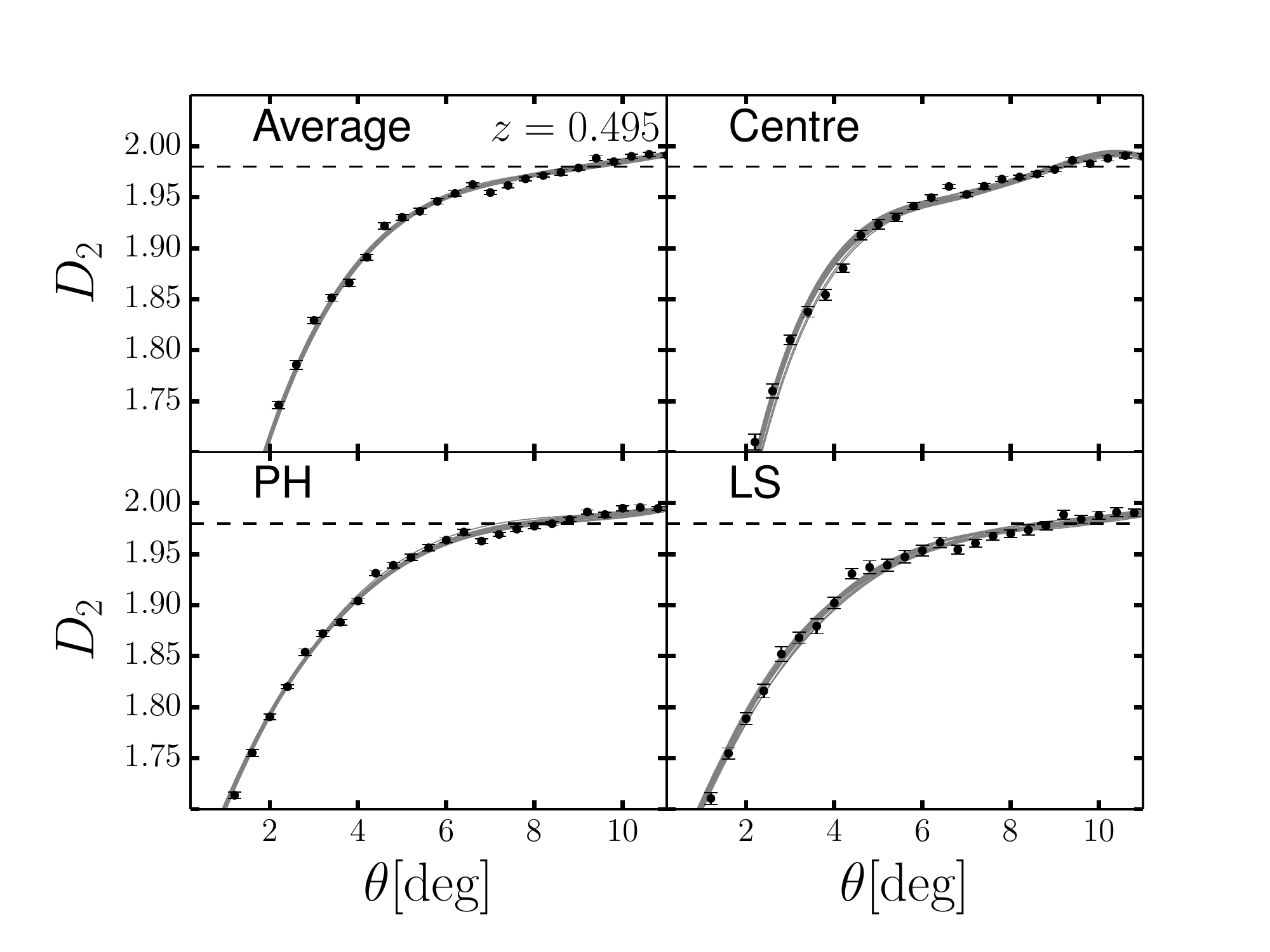}
%\epsfig{figure=mosaico-z0.495.eps,width=3.7truein,height=3.2truein}
%\vspace{-3.3cm}
\caption{Correlation dimension  for the four estimators, in the redshift bin $0.49 < z < 0.50$. The countinuous lines represent the polynomial fit performed for each catalogue.}
\label{fig1}
\end{figure}

In order to estimate the homogeneity scale, $\theta_{\rm h}$, for each one of the previous methods, we perform the following approach: we make a model-independent polynomial fit for each $D_{2}$ set, in each redshift slice (exemplified for one redshift slice in Fig. \ref{fig1}). Following previous analyses \citep{alonso14,alonso15,ntelis17}, we identify the scale of transition as the angle at which the fits of our estimator are within one per cent of the homogeneous limit $D_{2{\rm h}}$ given by Eq. (\ref{d2h}). Although arbitrary, the 1\%-criterion is widely used in the literature, and is justified given the sample noise. Given the values of $D_{2{\rm h}}$ we perform a bootstrap analysis ~\citep{efron83} on these values with 1000 realisations and we obtain the mean and error with 68\% c.l. for the $\theta_{\rm h}$ (Table \ref{table2}).

%--------------------------------------------------------------------------------------------------------------------------

%--------------------------------------------------------------------------------------------------------------------------

%--------------------------------------------------------------------------------------------------------------------------
\begin{figure}
\vspace{-0.5cm}
\includegraphics[scale=0.45]{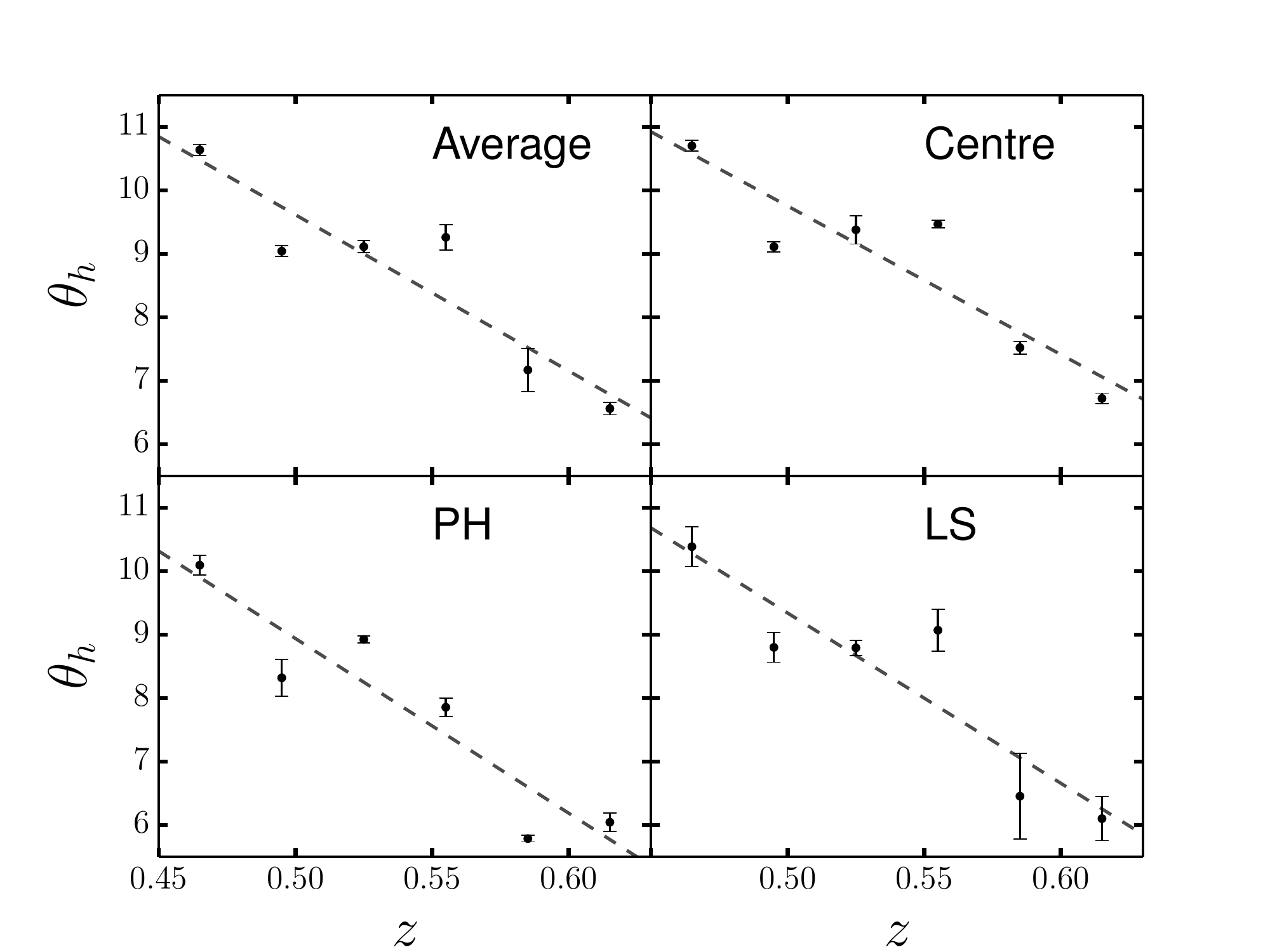}
%\epsfig{figure=mosaico-th-ajuste.eps,width=3.7truein,height=3.0truein}
%\vspace{-3.3cm}
\caption{Redshift evolution of the angular homogeneity scale for the four estimators. Data points are measurements in the redshift bins of Table \ref{table2}. The dashed line is a linear fit to the data points.}
\label{fig2}
\end{figure}
%----------------------------------------------------------------------------------------------------------------------
\vspace{3.0cm}

\begin{table}
\centering
\begin{tabular}{|c|c|c|c|c|c|c|c|c|}
\hline
$z$ & $\theta_{\rm Ave}$ & $\theta_{\rm Cen}$  & $\theta_{\rm PH}$  & $\theta_{\rm LS}$  \\
\hline
0.465 & $10.64 \pm 0.09$  & $10.70 \pm 0.08$ & $10.09 \pm 0.16$ & $10.39 \pm 0.31$ \\ \hline
0.495 & $9.04  \pm 0.09$  & $9.11  \pm 0.08$ & $8.32  \pm 0.29$ & $8.80  \pm 0.23$ \\ \hline
0.525 & $9.11  \pm 0.10$  & $9.38  \pm 0.22$ & $8.92  \pm 0.06$ & $8.79  \pm 0.12$ \\ \hline
0.555 & $9.26  \pm 0.20$  & $9.47  \pm 0.06$ & $7.86  \pm 0.14$ & $9.07  \pm 0.33$ \\ \hline
0.585 & $7.17  \pm 0.34$  & $7.52  \pm 0.10$ & $5.79  \pm 0.05$ & $6.46  \pm 0.67$ \\ \hline
0.615 & $6.56  \pm 0.10$  & $6.72  \pm 0.08$ & $6.04  \pm 0.14$ & $6.10  \pm 0.35$ \\ \hline
\end{tabular} 
\caption{Measurements of the angular homogeneity scale (degrees) for each redshift interval and estimator.}
\label{table2}
\end{table}
%----------------------------------------------------------------------------------------------------------------------

%--------------------------------------------------------------------------------------------------------------------------
\begin{table}
\begin{tabular}{|c|c|c|c|c|}
\hline
& $\alpha$ & $\beta$ & $\theta_{\rm h}(0.46) $ & $\theta_{\rm h}(0.62)$ \\
\hline
Average & 21.93 $\pm$ 2.81 & -24.62 $\pm$ 5.19 & 10.60 & 6.67 \\ \hline
Centre  & 21.46 $\pm$ 2.91 & -23.41 $\pm$ 5.37 & 10.69 & 6.95 \\ \hline
PH      & 22.71 $\pm$ 3.04 & -27.54 $\pm$ 5.61 & 10.04 & 5.63 \\ \hline
LS      & 22.76 $\pm$ 3.29 & -26.84 $\pm$ 6.07 & 10.41 & 6.12 \\ \hline
\end{tabular}
\caption{
For each estimator, the best-fits of $\alpha$ and $\beta$ in Eq. (\ref{lf}), and the predicted extrapolation of $\theta_{\rm h}$ 
at $z=0.46,0.62$. 
}
\label{table3}
\end{table}
\noindent

\section{Results}

Figure \ref{fig1} presents the fits of the correlation dimension for the four estimators, showing the crossing of the homogeneity threshold. We illustrate only the redshift slice $0.49 < z < 0.50$, since the results for the other slices are very similar. The corresponding numerical results for $\theta_{\rm h}$ and their errors are shown in Table \ref{table2}. We can observe that the four estimators produce similar $\theta_{\rm h}$ values.

Additionally, there is a clear correlation between $\theta_{\rm h}$ and $z$: for lower $z$, the transition angular scale increases, as illustrated in Fig. \ref{fig2}. This is the expected behaviour, since matter perturbations grow stronger in later epochs, so that the Universe should appear clumpier as the redshift decreases. To better visualize this correlation, we perform a linear fit, 
\be\label{lf}
\theta_{\rm h}(z) = \alpha + \beta z,
\ee 
and calculate the parameters $\alpha$ and $\beta$ for each estimator. The results are shown in Table \ref{table3}. One can see the four estimators show the same trend and the maximum dispersion of the slopes is $\sim 15\%$.

In order to compare our results with previous model-dependent analyses, we convert the $\theta_{\rm h}$ measurements in Table \ref{table3} into the corresponding physical distance, $r_{\rm h}(z) = D_A(z) \theta_{\rm h}(z)$. For the comparison, we consider two redshifts, $z = 0.46$ and $z = 0.62$, and use the latest best-fit $\Lambda$CDM cosmology from the Planck Collaboration, with $\Omega_m=0.308$ and $h=0.678$ \citep{planck16}. We obtain a spatial homogeneity scale 
\be\label{rh} 
218 \leq\! r_{\rm h}(0.46)\! \leq 232\, \mathrm{Mpc} ,~121 \leq \!r_{\rm h}(0.62)\! \leq 151\, \mathrm{Mpc} ,
\ee 
considering the lowest and highest $\theta_{\rm h}$ values in Table \ref{table3}. These results are compatible (within $2\sigma$) with the estimates in \cite{ntelis17} for the same DR12 LRG catalogue, where $r_{\rm h}(\bar{z} = 0.46) = 185.1 \pm 18.6 \, \mathrm{Mpc}$, and $r_{\rm h}(\bar{z} = 0.62) = 161.7 \pm 5.8 \, \mathrm{Mpc}$, based on a $\Lambda$CDM model.

Our results are also compatible with~\cite{pandey15}, which used the DR12 Main Galaxy Sample to find $r_{\rm h} \simeq 206 \, \mathrm{Mpc}$. In addition, they are consistent with the upper limit estimate of $r_{\rm h} \simeq 383 \, \mathrm{Mpc}$ \citep{yadav10}. We emphasize that these analyses were performed in a model-dependent framework ($\Lambda$CDM) to convert redshifts and angles into distances, whereas our analysis only requires angular information of the galaxy distribution. Therefore, our results are consistent with the standard cosmological scenario even without assuming a particular FLRW model.

\section{Conclusion}

The assumption of large-scale spatial homogeneity and isotropy is at the root of modern cosmology. Although spatial isotropy has been tested using different methods and probes, the homogeneity hypothesis is much more difficult to probe. Tests that are based on source counts in spheres or caps can be classified as consistency tests of the Cosmological Principle, since they do not probe inside the past lightcone. Tests based on a length scale $r_{\rm h}$ must further assume a fiducial FRLW model, in order to relate redshifts and angles to distances. In this {\it Letter}, we estimated the cosmological angular homogeneity scale, following an approach that avoids the need to assume a fiducial cosmological model, and that is based only on observable quantities. We used a sample of 159,391 LRG provided by BOSS DR12. To perform our measurements, we  divided the sample into 6 redshift bins in the range $0.46\leq z\leq 0.62$, which provides at least 18,800 galaxies per bin. Our analysis was carried out using four different estimators to compute the correlation dimension, which showed a well agreement between them (see Fig. \ref{fig1} and Table \ref{table2}). 

By using non-contiguous redshift slices, we suppress correlations between the slices, which otherwise could bias the results. The thinness of the redshift bins, $\Delta z=0.01$, means  that we do not falsely introduce homogenization by projecting sources that have large radial separation into the same spherical cap. In addition, evolution in these bins can safely be ignored. Redshift-space distortions will move galaxies into and out of redshift bins, but the effect should average out, given the high number of galaxies.

Thanks to the depth of the data sample, we were also able to investigate the redshift evolution of the angular homogeneity scale, shown in Fig. \ref{fig2}. We found a clear correlation between $\theta_{\rm h}$ and $z$, in which the lower the redshift the larger the transition angular scale. We applied a simple linear fit to $\theta_{\rm h}(z)$ and calculated the expected transition scales at $z = 0.4$ and $z = 0.6$, shown in Eq. (\ref{rh}). We compared our measurements at these redshifts with previous model-dependent analyses of the same dataset, by transforming $\theta_{\rm h}$ into $r_h$. Even without assuming a fiducial cosmological model, our results are in good agreement with transition homogeneity scales obtained in~\cite{pandey15, ntelis17}, as well as with the theoretical upper limit prediction for the standard $\Lambda$CDM cosmology~\citep{yadav10}. 

In summary, we showed that the hypothesis of large-scale homogeneity in the LRG distribution seems to be in good concordance with the current cosmological scenario. The method discussed here, which is a spectroscopic and tomographic extension of the method originally proposed in~\cite{alonso14}, can be applied to current and upcoming surveys, such as SDSS-IV (eBOSS)~\citep{eboss16}, J-PAS~\citep{jpas14}, Euclid~\citep{euclid16}, LSST~\citep{lsst09}, and SKA~\citep{ska15}. \\

\section*{Acknowledgments}

We acknowledge the use of the Sloan Sky Digital Survey data \citep{york00}. RM thanks David Alonso for useful discussions. RSG, GCC, CAPB and JCC thank CNPq for support. AB acknowledges a PVE project from CAPES (Science without Borders program), no. 88881.064966/2014-01. JSA acknowledges support from CNPq (Grants no. 310790/2014-0 and 400471/2014-0) and FAPERJ (grant no. 204282). CAPB and RM acknowledge support from the South African SKA Project. RM is also supported by the UK STFC, grant no. ST/N000668/1.

\end{document}